\documentclass[acmsmall]{acmart}
\AtBeginDocument{%
  }

\setcopyright{acmlicensed}

\acmJournal{JACM}
\acmVolume{0}
\acmNumber{0}
\acmArticle{0}
\acmMonth{0}

\usepackage{subfigure}

\begin{document}

\title{Cloud Native System for LLM Inference Serving}

\author{Minxian Xu, Junhan Liao, Jingfeng Wu, Yiyuan He, Kejiang Ye}
	\email{{mx.xu, jh.liao2, jf.wu2, yy.he2, kj.ye}@siat.ac.cn}
        \authornote{Corresponding author}
	\affiliation{%
		\institution{Shenzhen Institute of Advanced Technology, Chinese Academy of Sciences; University of Chinese Academy of Sciences}
		\country{China}
	}
	
	\author{Chengzhong Xu}
	\email{czxu@um.edu.mo}
	\affiliation{%
		\institution{State Key Lab of IOTSC, University of Macau}
		\country{China}
	}

\renewcommand{\shortauthors}{Xu et al.}

\begin{abstract}
Large Language Models (LLMs) are revolutionizing numerous industries, but their substantial computational demands create challenges for efficient deployment, particularly in cloud environments. Traditional approaches to inference serving often struggle with resource inefficiencies, leading to high operational costs, latency issues, and limited scalability. This article explores how Cloud Native technologies—such as containerization, microservices, and dynamic scheduling—can fundamentally improve LLM inference serving. By leveraging these technologies, we demonstrate how a Cloud Native system enables more efficient resource allocation, reduces latency, and enhances throughput in high-demand scenarios. Through real-world evaluations using Kubernetes-based autoscaling, we show that Cloud Native architectures can dynamically adapt to workload fluctuations, mitigating performance bottlenecks while optimizing LLM inference serving performance. This discussion provides a broader perspective on how Cloud Native frameworks could reshape the future of scalable LLM inference serving, offering key insights for researchers, practitioners, and industry leaders in cloud computing and artificial intelligence.
\end{abstract}

	\ccsdesc[500]{Networks~Cloud computing}
	\ccsdesc[500]{Computing methodologies~Distributed algorithms}
\keywords{Cloud Native, LLM, Inference Serving, Resource Management}

\maketitle

\section{Introduction}

Since the release of GPT-3.5 \cite{GPT} in 2022, LLMs have gained widespread attention and are now applied in various domains, including healthcare, law, and education. Rapid advancement of LLMs has led to the development of numerous new models, such as the GPT series (GPT-1, GPT-2, GPT-3, etc.), the LLaMA family (LLaMA2 \cite{LLaMA2}, LongChat \cite{LongChat}), BLOOM \cite{BLOOM}, FALCON \cite{FALCON}, GLM \cite{GLM}, Mistral \cite{Mistral}, PaLM \cite{PaLM}, and DeepSeek \cite{DeepSeek}. These models have attracted strong interest from both academia and industry, fueling innovation in artificial intelligence and natural language processing. 
\begin{figure}
    \centering
    \includegraphics[width=\textwidth]{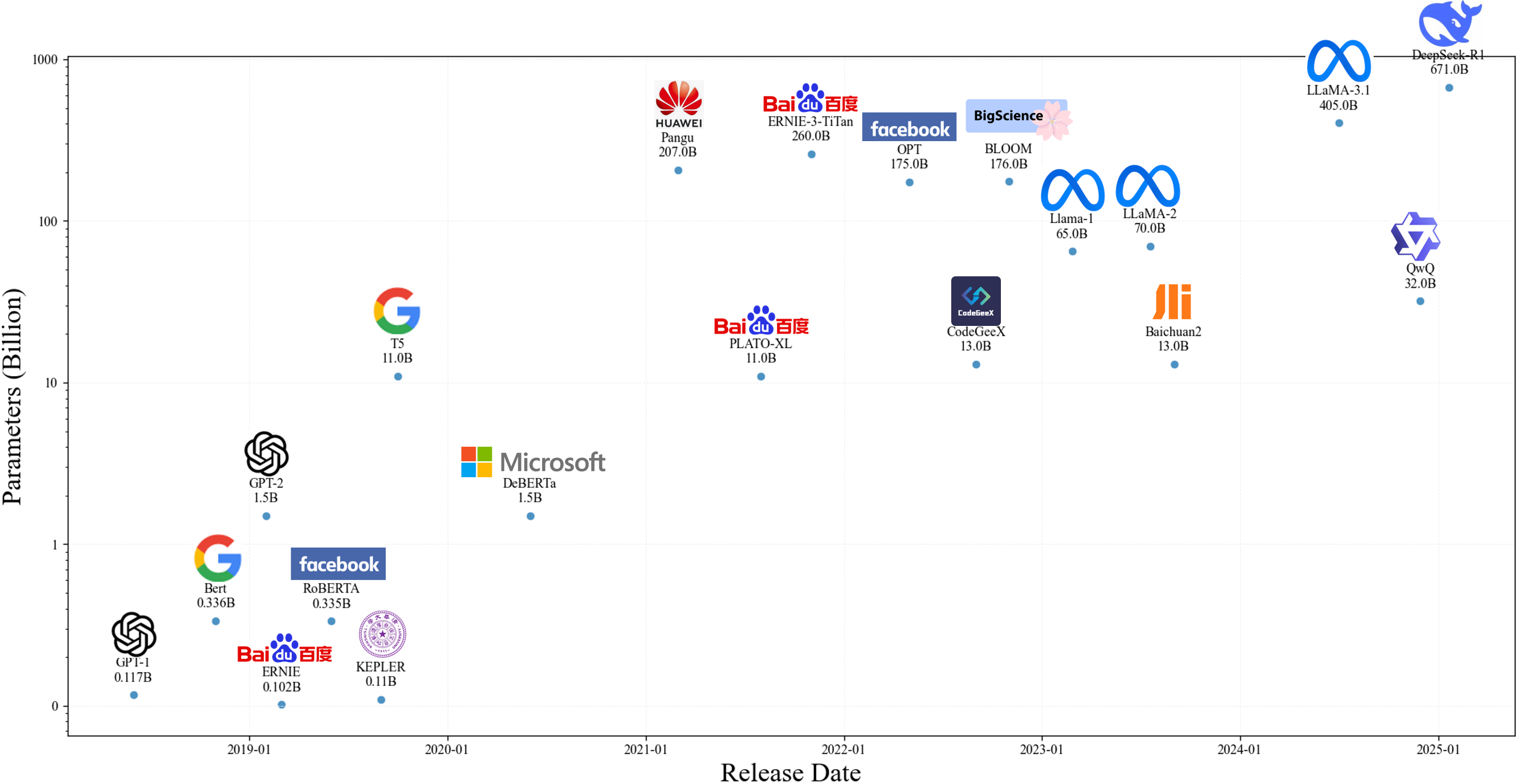}
    \caption{The Development of Parameters in Current Mainstream Open-source LLMs.}
    \label{fig:LLM_Parameter_Develop}
\end{figure}
However, as these models grow in scale and complexity, their computational demands have increased exponentially. Figure \ref{fig:LLM_Parameter_Develop} illustrates the evolution of LLM parameters, revealing a trend to increase the size of the model over time. Training state-of-the-art models such as GPT-4 requires at least 25,000 NVIDIA A100 GPUs running continuously for 90 to 100 days, while deploying these models for real-time inference demands at least 128 A100 GPUs to maintain service performance. With the widespread adoption of LLMs, inference has become the dominant cost factor in cloud-based machine learning, accounting for up to 90\% of total expenditures.

\textit{LLM inference serving}, the process of deploying trained models to generate real-time predictions, is computationally intensive due to frequent auto-regressive computations, memory access, and storage operations. As a result, inference performance is highly dependent on GPU resources, directly affecting both service efficiency and user experience. Furthermore, increasing user expectations demand lower latency and greater stability in LLM responses. Two key performance metrics define inference efficiency: latency and throughput. Latency is primarily determined by Time to First Token (TTFT), Time Per Output Token (TPOT), and the total response token count, while throughput reflects the number of requests processed per second. Traditional LLM inference services often suffer from inefficiencies. To improve throughput, techniques such as PagedAttention \cite{PagedAttention}, Continuous Batching \cite{ContinuousBatching} have been developed and integrated into inference systems. Currently, methods such as Llumnix \cite{Llumnix} and DistServe \cite{DistServe}, which leverage request migration strategies, have been proposed to reduce end-to-end latency, queueing delays, and TTFT. However, these optimizations introduce trade-offs: reducing latency often comes at the cost of lower throughput, while increasing throughput may lead to significantly higher per-request latency, making it challenging to achieve an optimal balance between the two.

Despite advances in LLM technology, cloud-based inference deployment remains inefficient, leading to considerable resource under-utilization and scalability challenges. Most LLMs rely on Transformer architectures with self-attention mechanisms and are typically deployed as large, monolithic models. Their sheer size (e.g., GPT-4 with 1.8 trillion parameters) results in high overhead costs related to initialization and replication. In high-concurrency, low-latency environments, cloud providers frequently over-provision model replicas to ensure service quality, but this approach exacerbates resource inefficiencies. For example, GPU clusters in data centers often operate at a low utilization 30\%. Although multiple replicas of models are deployed to handle peak demand, more than 70\% of the allocated resources remain idle outside of peak usage periods. This imbalance in resource allocation significantly limits the scalability and commercial viability of LLM services.

Addressing these inefficiencies requires innovative approaches to resource scheduling and deployment strategies. One promising direction is Cloud Native computing, an architectural paradigm designed to maximize the scalability, flexibility, and efficiency of applications in cloud environments. Cloud Native systems leverage technologies such as containerization, microservices, and dynamic orchestration to enable more efficient resource management. Unlike traditional monolithic deployments, Cloud Native architectures break down applications into modular components that can be independently managed, scaled, and optimized based on real-time demand. By applying these principles to LLM inference serving, Cloud Native frameworks can significantly enhance resource utilization, reduce latency, and improve throughput. This article explores how Cloud Native technologies can transform LLM deployment, offering a scalable and cost-effective solution for cloud-based AI services.

The \textbf{key insights} of this article are as follows: 
\begin{itemize}
\item {Large-scale LLM inference serving faces fundamental challenges in resource utilization, often leading to inefficiencies in cloud environments.}
\item {Achieving optimal inference performance is difficult due to workload variability, dynamic resource demands, and imbalances in allocation strategies.}
\item {Cloud Native technologies provide a structured approach to improving LLM inference serving by enhancing scalability, efficiency, and adaptability.}
\end{itemize}

\section{Why Is LLM Inference Serving Hard to Optimize?}

Can resource utilization efficiency be improved while maintaining the Service Level Objective (SLO) of inference services? This question remains a critical challenge for LLM research teams worldwide. Several fundamental factors contribute to the difficulty of optimizing inference performance:

\emph{\textbf{Uncertainty in Input and Output Length}}: LLM inference services handle diverse inputs, with highly variable and unpredictable output lengths. Static resource configurations struggle to accommodate these variations, making it difficult to efficiently allocate computational resources and predict system load.

\emph{\textbf{Imbalance in Inference Stages}}: The inference process consists of multiple stages, such as prefill and decode, each with different resource demands. Here, the prefill stage refers to the process of generating the first token by LLMs, while the decode stage refers to the process of generating each subsequent token after the first. The prefill stage may require intensive computation, while the decode stage is more memory-bound. This imbalance can lead to inefficient resource usage, with certain hardware components underutilized while others are overloaded.

\emph{\textbf{Inefficiencies in Traditional Resource Allocation}}: Static resource allocation methods fail to adapt to dynamically changing workloads and varying SLOs. Pre-reserving resources and using fixed configurations often result in over-provisioning (leading to wasted resources) or under-provisioning (causing performance degradation).

\emph{\textbf{Lack of Fine-Grained System Adaptability}}: Many inference systems are designed with coarse-grained modular architectures that lack the flexibility to adjust resource allocation dynamically. This rigidity makes it difficult to respond efficiently to workload fluctuations and high-concurrency scenarios.

\emph{\textbf{Resource Competition and Scheduling Bottlenecks}}: In high-concurrency environments, multiple requests compete for limited computing resources, increasing service latency. Static scheduling strategies struggle to handle this contention, further exacerbating performance delays and inconsistencies.

\emph{\textbf{Hardware and Resource Imbalances}}: Mismatches between hardware capabilities—such as GPU compute power, memory bandwidth, and network throughput—often create bottlenecks in specific inference tasks. These imbalances lead to suboptimal system performance and reduced overall efficiency.

\emph{\textbf{Challenges in Managing Burst Workloads}}: While inference systems are designed with anticipated demand in mind, they often fail to handle unexpected traffic spikes effectively. This limitation results in performance degradation and increased response time during peak loads.

\emph{\textbf{Trade-Off Between Latency and Throughput}}: A persistent challenge in inference serving is balancing latency and throughput. Reducing latency can decrease overall system throughput, while optimizing for throughput can introduce excessive delays for individual requests. Finding an optimal balance remains a complex problem in large-scale LLM deployment.

Given the challenges of unpredictable workloads in LLM inference and the coarse granularity of existing system designs, three key issues commonly emerge: inaccurate resource profiling, rigid scheduling and orchestration, and imbalanced model deployment.

\subsection{Inaccurate Resource Profiling}

LLM inference tasks exhibit complex and unpredictable load patterns, making it difficult to develop precise resource profiles. These profiles are critical for optimizing operational efficiency and resource utilization, especially in distributed and multi-node environments. However, several key challenges hinder accurate profiling:

\textbf{Limited Adaptability to Dynamic Hardware Conditions}: LLMs run on diverse hardware platforms (e.g., GPUs and TPUs), yet real-time monitoring of hardware variations remains inadequate. Factors such as GPU thermal throttling, scheduling conflicts, and background processes can degrade performance. When resource profiles fail to capture these fluctuations, scheduling and optimization strategies become less effective, leading to inefficiencies in model execution.

\textbf{Inaccurate Task Load Prediction}: The workload of LLMs depends on variables such as input size and model complexity. Without precise load predictions, resource scheduling becomes suboptimal, affecting computational efficiency and response time. In multi-GPU or multi-node environments, uneven load distribution can increase latency and degrade overall system performance.


\textbf{Variability Across Inference Stages}: Inference consists of distinct stages, such as prefill (model loading and input processing) and decode (token generation). These stages impose different demands—prefill is computation-heavy, while decode relies more on memory and storage bandwidth. Additional complexities, such as frequent storage updates during token generation and shared context management in multi-task scenarios, make precise resource modeling even more challenging.

These challenges contribute to bottlenecks during peak workloads, ultimately affecting inference service quality, scalability, and responsiveness, which requires \textit{accurate resource profiling} for LLM inference task.

\subsection{Rigid Scheduling and Orchestration}

In large-scale distributed environments, orchestration and scheduling are critical for managing resources efficiently. However, existing approaches often introduce excessive complexity and operational burdens, leading to suboptimal performance. The key challenges include:

\textbf{High Complexity of Scheduling Algorithms.} Many scheduling algorithms focus on global optimization to maximize resource utilization. However, in practice, these approaches introduce significant computational and communication overhead. Factors such as inter-node communication delays and data locality must be accounted for, increasing algorithmic complexity and slowing down decision-making.


\textbf{Limited Real-Time Adaptability.} LLM inference requires low-latency, real-time scheduling, but many traditional systems rely on periodic scheduling methods that struggle to adapt to sudden workload fluctuations. This lag in response time can degrade service quality, particularly in high-demand scenarios where immediate resource adjustments are necessary.

\textbf{Coarse-Grained Model Management.} Cloud-based LLM inference often suffers from coarse-grained resource allocation, where entire models are treated as monolithic units. This approach results in high computational and storage costs, making model loading, migration, and replication inefficient. As a result, during peak periods, the system struggles to dynamically reallocate resources in response to changing demand, leading to service delays and underutilized infrastructure.

These limitations highlight the need for more \textit{flexible, fine-grained scheduling and orchestration} strategies that can dynamically respond to workload changes while maintaining efficient resource utilization. 

\subsection{Imbalanced model deployment}

In large-scale deployments, efficient resource allocation is essential for ensuring optimal performance and cost-effectiveness. However, uneven distribution of computational resources often leads to inefficiencies, where some nodes experience excessive workloads while others remain underutilized. Key challenges include:

\textbf{Insufficient Data Parallelism.} LLM inference often relies on data parallelism to distribute workloads across multiple computing units. However, this distribution is frequently uneven due to factors such as data bias and workload imbalance. Some nodes may process more data or perform more complex computations than others, creating bottlenecks while leaving other resources idle.

\textbf{Uneven Computational Demand Across Model Components.} Different layers of LLM architectures—such as those in Transformer models—have varying computational requirements. Current deployment strategies do not effectively account for these differences, leading to disproportionate workloads on certain nodes, which limits overall system efficiency.

\textbf{Suboptimal Hardware Resource Utilization.} LLM inference typically runs on a mix of hardware resources, including GPUs and TPUs, each with varying computational power, memory bandwidth, and network capabilities. Existing deployment strategies often fail to consider these hardware discrepancies, leading to overloaded nodes and underutilized resources. Additionally, most deployments rely on static, pre-configured resource allocations, reserving excess capacity for peak demand periods. This approach results in low resource utilization during off-peak times and limits the system’s ability to adapt to traffic fluctuations.

Addressing these challenges requires \textit{adaptive, real-time resource allocation} strategies that leverage dynamic monitoring and predictive scaling to balance workloads effectively. By enabling more flexible and intelligent deployment mechanisms, cloud-based LLM inference services can significantly improve efficiency, reduce costs, and enhance scalability.

\section{Cloud Native For LLM}
Based on the identified challenges, we propose a solution that leverages cloud native technology for LLM applications. A cloud native system is an architectural paradigm designed to optimize and enhance the deployment, management, and scalability of large-scale applications through techniques such as \textit{containerization, microservices, and dynamic scheduling}. In traditional cloud computing architectures, applications tend to be monolithic with static resource allocation and management. In contrast, cloud native approaches advocate decomposing applications into small, independent modules (also called microservices) that can operate, be deployed, and scale independently \cite{cloudNativeAdvantages}. This approach offers flexible scalability, automated resource management, and efficient deployment—attributes that are particularly advantageous for running LLMs efficiently.

A comprehensive review of both domestic and international literature reveals that existing LLM inference services have yet to fully exploit the potential of cloud native systems, lacking a precise, fine-grained, and dynamic prototype framework. To overcome the performance optimization challenges encountered in current LLM inference services, we suggest integrating the efficient orchestration and resource management strategies \cite{ContainerOrchestration} inherent in cloud native systems into task profiling, module expansion, and model deployment, thereby constructing a cloud native based inference service platform. Figure \ref{fig:cloudnative-llm} illustrates the proposed cloud native LLM inference service architecture, systematically organized as a multi-tiered framework to address the limitations identified in existing approaches. The top tier consists of various LLM inference workloads that will be transmitted to the cloud native platform from different LLMs.  The third and fourth tiers depict the AI infrastructure used to build the cloud native platform (e.g., public and private clouds) and the types of hardware it employs (e.g., GPU, TPU and NPU). The second tier presents the key components within the cloud native platform  to support LLMs inference serving, which can be broadly categorized into six modules as below:
\begin{figure}
    \centering
    \includegraphics[width=\textwidth]{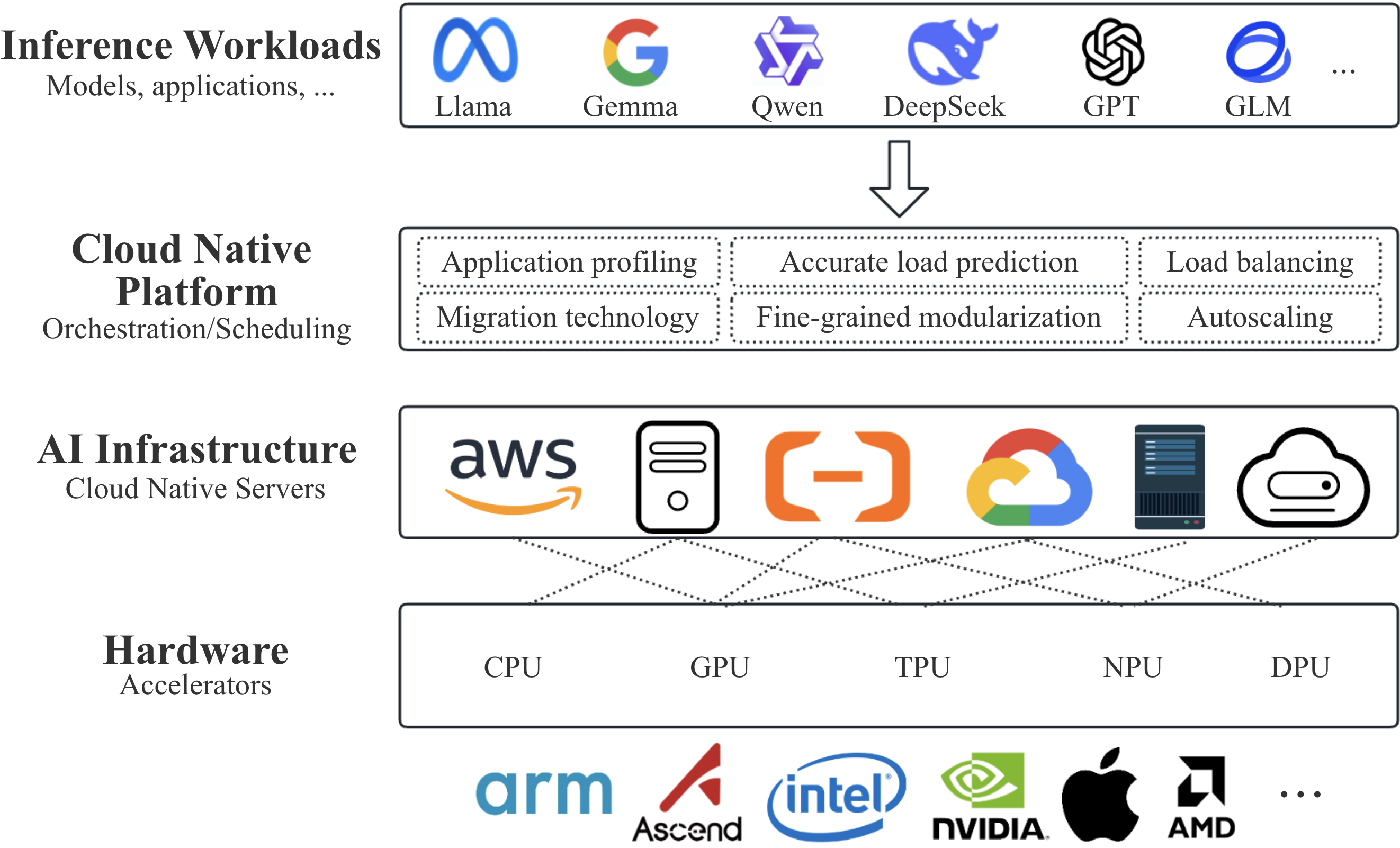}
    \caption{Cloud Native System Support Framework for LLM Inference Serving.}
    \label{fig:cloudnative-llm}
\end{figure}

\textbf{Load balancing} is one of the critical factors for enhancing the performance of LLM inference services. In distributed systems, the load associated with inference tasks is inherently dynamic, particularly under high request volumes or when input data varies significantly, leading to potential load imbalances. Without an effective load balancing mechanism, some nodes may become overburdened while others remain underutilized, resulting in resource wastage and increased response time. Intelligent load balancing in cloud native environments dynamically adjusts task allocation based on real-time resource usage metrics—such as computing power, memory, and bandwidth—across nodes. For instance, Kubernetes-based \cite{Kubernetes} service meshes (e.g., Istio) can automatically redistribute traffic among service instances in response to real-time demand, ensuring that each request is directed to a node with lower load, thereby mitigating bottlenecks.

\textbf{Autoscaling} represents another key technology for optimizing LLM inference services. In cloud native settings, it is crucial for the inference service to dynamically adjust computing resources in response to fluctuating loads, especially during peak periods. Autoscaling mechanisms automatically increase or decrease resource capacity based on real-time load data, ensuring optimal resource utilization and high system availability. For example, Kubernetes’ Horizontal Pod Autoscaler (HPA) adjusts the number of Pods (independent microservices) based on CPU or memory usage; it scales out the number of instances when faced with a surge in inference requests, and scales them back when the load diminishes, thus balancing performance and cost-efficiency.

\textbf{Migration technology} also plays a vital role in optimizing the performance of LLM inference services, particularly under conditions of resource demand imbalance. This technology facilitates the transfer of tasks across different nodes, ensuring better resource utilization and preventing overload on individual nodes. In high-performance GPU environments, for example, LLM inference tasks may overload certain GPUs, thereby reducing execution efficiency. Through transparent migration enabled by cloud native systems, tasks can be relocated to less burdened nodes or GPUs, achieving more efficient resource use and lower latency without manual intervention.

\textbf{Accurate load prediction} is another essential strategy for optimizing LLM inference performance. By forecasting future load conditions through the analysis of historical data with machine learning algorithms, systems can preemptively allocate resources and schedule tasks. For instance, integrating time series prediction models enables the anticipation of high-load periods, allowing proactive resource scaling and adjustments. This approach not only prevents service interruptions due to insufficient resources but also minimizes the waste associated with over-provisioning, thereby optimizing resource allocation.

\textbf{Application profiling} is indispensable for performance optimization in LLM inference services. In a cloud native framework, application profiling involves comprehensive monitoring and modeling of an application’s performance, resource requirements, and behavioral patterns, which helps in understanding the specific demands of each inference request. By collecting detailed operational data—such as processing latency, resource usage, and failure rates—profiling provides a robust basis for decisions regarding load balancing, autoscaling, and migration.

\textbf{Fine-grained modularization}. Furthermore, the modular design inherent in cloud native systems facilitates more granular monitoring and control, making it feasible to optimize individual components of the model. Unlike traditional monolithic deployments, where the entire model inference pipeline is treated as a single unit, fine-grained modularization decomposes the inference process into multiple independent functional components. This approach enables selective optimization, independent scaling, and precise fault isolation, ultimately improving resource utilization and system resilience.

Through these techniques, cloud native systems can significantly improve service efficiency by enhancing resource utilization, reducing operating costs, and boosting both model performance and stability. Ultimately, cloud native not only provides architectural flexibility and scalability but also offers robust support for the efficient operation of LLM applications in large-scale computing environments.
\section{LLM Serving Performance Improvement with Cloud Native}

In this section, we introduce our experimental framework and findings aimed at evaluating the performance enhancements achieved by applying cloud native techniques to LLM serving. Our objective is to demonstrate that decomposing LLMs into microservices and employing dynamic autoscaling can effectively mitigate bottlenecks in Transformer layers, thereby reducing inference latency and increasing overall throughput.

\subsection{Experimental Settings}

Traditional deployments treat the entire model as a monolithic unit with uniform resource allocation. In contrast, cloud native systems permit granular decomposition, such as isolating each Transformer layer of an LLM as an independent microservice for targeted optimization. To validate this approach, we conducted a series of experiments focusing on load bottlenecks and the impact of autoscaling on a bottleneck Transformer layer.

Our experimental environment comprised a Kubernetes cluster with three nodes, each equipped with an NVIDIA A100-80GB GPU interconnected via NVLink. We employed the LLaMA-2-13B model, decomposing each of its Transformer layers into separate microservices. Inter-service communication was managed via gRPC\footnote{https://grpc.io/} and orchestrated using the Istio\footnote{https://istio.io/} service mesh. Monitoring was performed with Prometheus\footnote{https://prometheus.io/} and Grafana\footnote{https://grafana.com/}, which collected data on GPU utilization, memory bandwidth, and per-layer forward propagation times at a 100ms sampling interval. To simulate diverse load conditions, Locust\footnote{https://locust.io/} toolkit generated requests with input lengths varying from 50 to 2048 tokens. 

Based on the latency distribution, we identified a specific Transformer layer with a pronounced right-skewed latency profile as the primary bottleneck. To assess the effectiveness of cloud native autoscaling (CN autoscaling), we first established a baseline without autoscaling (w/o autoscaling) which disables the Kubernetes HPA, recording key performance metrics such as average latency, peak latency, and latency jitter under varied workloads. We then deployed HPA for the microservice hosting the bottleneck layer by setting target GPU utilization and custom latency thresholds to trigger horizontal scaling.
\begin{figure}
    \centering
    \includegraphics[width=0.75\textwidth]{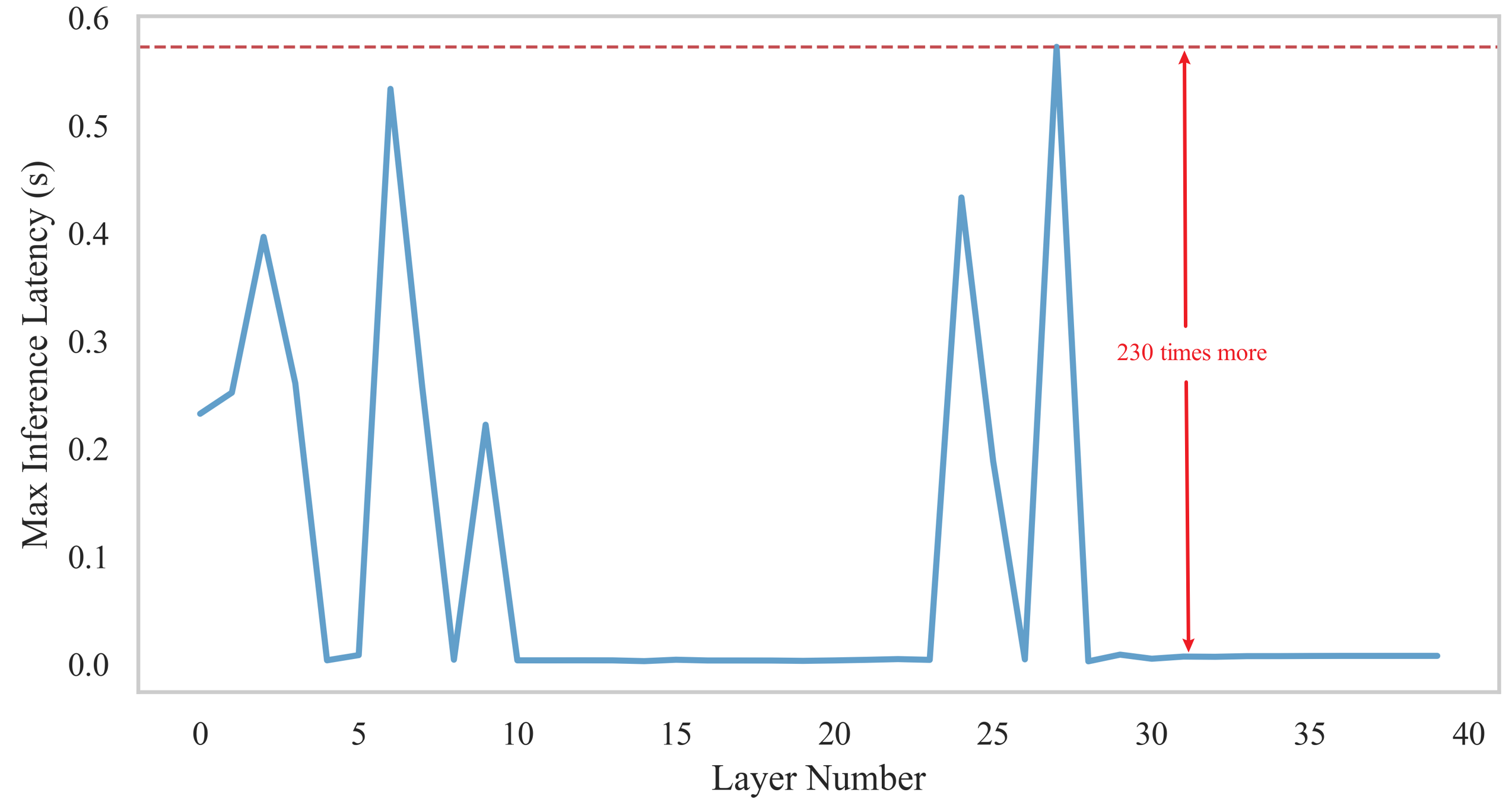}
    \caption{Maximum Inference Latency at Different Transformer Layers.}
    \label{fig:Layer_bottleneck}
\end{figure}

\subsection{Observations and Results}

The experimental results yielded several notable observations:

\begin{figure}[]
\centering
\subfigure[Inference latency comparison.]{\label{fig:latency}
\includegraphics[width=0.4\textwidth]{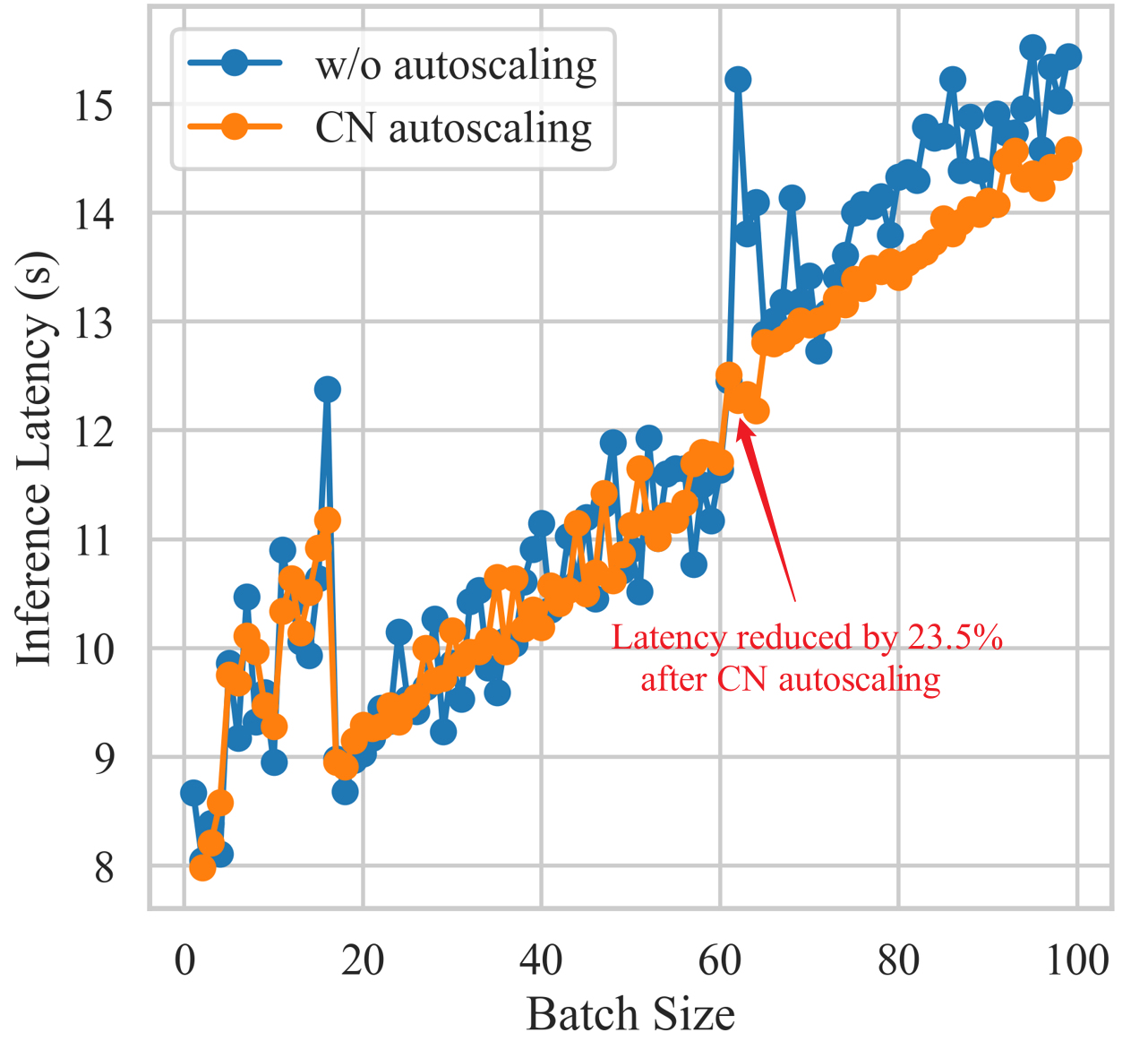}}
\subfigure[Throughput comparison.]{\label{fig:throughput}
\includegraphics[width=0.4\textwidth]{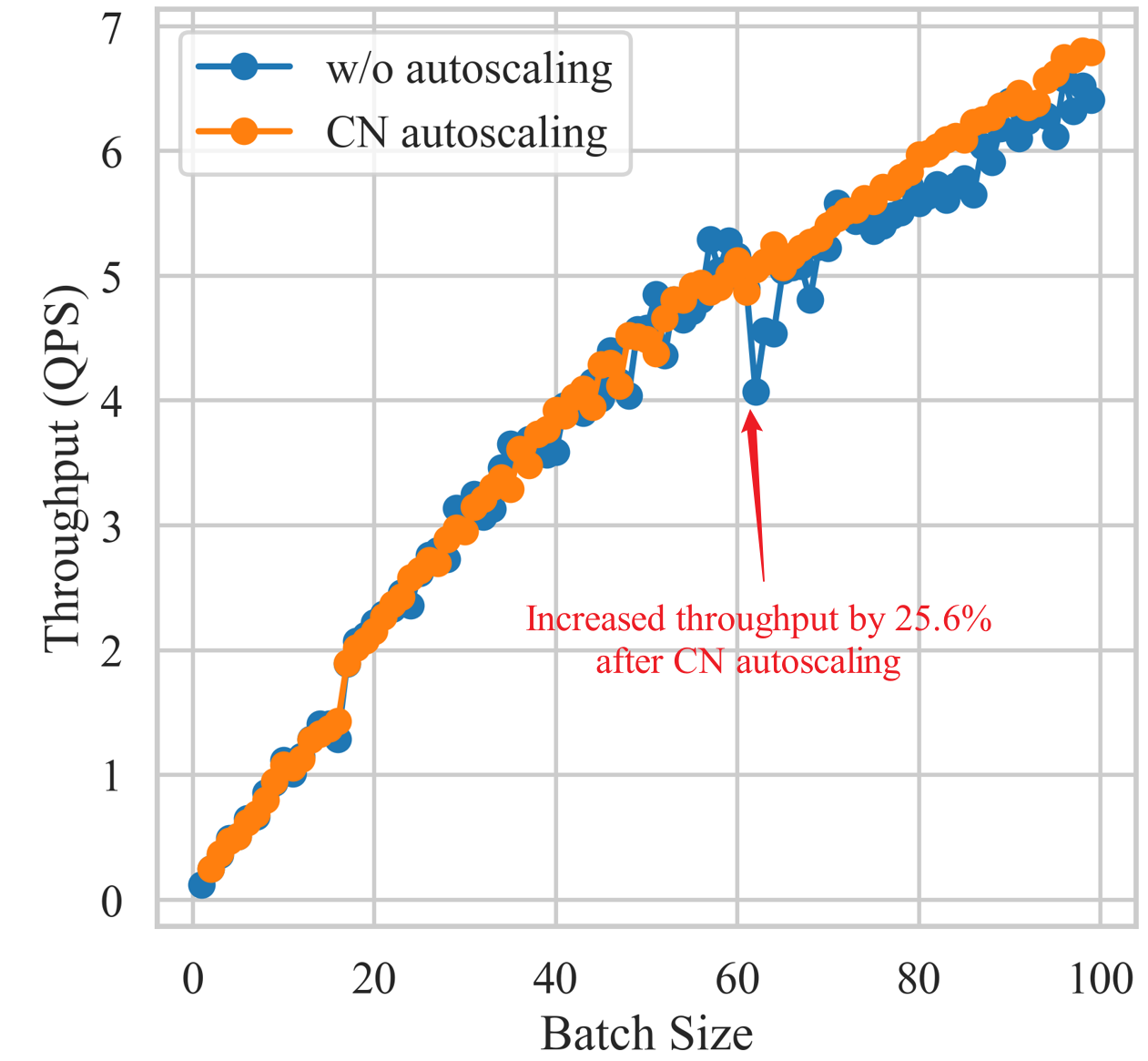}}
\caption{Performance Improvement with Autoscaling in Cloud Native.}
\label{fig:scale_layer}
\end{figure}

\textbf{Bottlenecks Identification:}
Figure \ref{fig:Layer_bottleneck} displays the maximum inference latency observed across 40 Transformer layers, highlighting distinct performance bottlenecks—particularly in layers exhibiting greater computational or memory overhead. Our analysis of 40 Transformer layers revealed that several layers, especially Layer 27, whose maximum latency was more than 230 times that of Layer 30, exhibited significant latency jitter and extreme peak latencies under high load. While low-load conditions produced relatively uniform latencies across layers, increased concurrency led to dramatic latency surges in bottleneck layers, indicating higher susceptibility to saturation from computational, memory, or scheduling constraints.

\textbf{Autoscaling Effects:}
Figure \ref{fig:scale_layer} presents a comparison of inference throughput and latency under different batch size settings between the default w/o autoscaling and CN autoscaling. Comparative tests before and after enabling CN autoscaling demonstrated that CN autoscaling substantially improved performance. As shown in Fig. \ref{fig:latency}, with a batch size of 62, the inference latency of the targeted bottleneck layer decreased from 15.23 seconds to 12.28 seconds, with a notable reduction in peak latency and mitigation of the long-tail effect as evidenced by a leftward shift in terms of latency. As shown in Fig. \ref{fig:throughput}, when the batch size was 62, under the same request load, the overall system throughput increased from 4.07 queries per second (QPS) to 5.05 QPS. These improvements are attributed to the dynamic load balancing achieved through horizontal scaling, which redistributed the computational load across additional pods.

To summarize, our experiments confirm that decomposing LLMs into fine-grained microservices and employing cloud native autoscaling via Kubernetes HPA effectively alleviates performance bottlenecks in Transformer layers. This strategy not only reduces the inference latency of critical layers but also enhances the overall throughput and stability of the LLM serving system. The results underscore the potential of cloud native architectures for optimizing large language model inference in dynamic and high-load environments.

\section{Conclusions}

In this paper, we demonstrated that a cloud native platform is highly effective in optimizing LLM inference. By decomposing monolithic LLM architectures into fine-grained microservices, our approach enables targeted performance improvements through dynamic load balancing and resource management. In particular, the application of cloud native autoscaling to identified bottleneck modules significantly reduced inference latency and enhanced throughput. The experimental results validate that cloud native techniques not only mitigate layer-specific performance bottlenecks but also facilitate efficient resource utilization under varying load conditions, underscoring their promise for LLM inference serving.

\begin{acks}
This work is supported by Guangdong Basic and Applied Basic Research Foundation (No. 2024A1515010251, 2023B1515130002), Guangdong Special Support Plan (No. 2021TQ06X990), and Shenzhen Basic Research Program under grants JCYJ20220818101610023 and JCYJ20240809180935001.
\end{acks}

\bibliographystyle{ACM-Reference-Format}
\bibliography{references}

\end{document}